# A fast quantum mechanical algorithm for estimating the median

Lov K. Grover
3C-404A Bell Labs
600 Mountain Avenue
Murray Hill NJ 07974
*lkg@mhcnet.att.com*

## Summary

Consider the problem of estimating the median of $N$ items to a precision $\varepsilon$, i.e. the estimate $\mu$ should be such that, with a large probability, the number of items with values smaller than $\mu$ is less than $\frac{N}{2}(1+|\varepsilon|)$ and those with values greater than $\mu$ is also less than $\frac{N}{2}(1+|\varepsilon|)$. Any classical algorithm to do this will need at least $\Omega\left(\frac{1}{\varepsilon^2}\right)$ samples. Quantum mechanical systems can simultaneously carry out multiple computations due to their wave like properties. This paper gives an $O\left(\frac{1}{|\varepsilon|}\right)$ step algorithm for the above problem.

## 1 Introduction

**1.1 Statistical Problems** Statistical problems involve processing large amounts of data. By their nature quantum mechanical systems can simultaneously carry out multiple computations and if adjusted properly, the overall computation can be significantly faster. [Shor94] used the fact that the periodicity of a function can be obtained rapidly by a quantum mechanical algorithm to develop a logarithmic time integer factorization algorithm. [Gro96] presented an $O(\sqrt{N})$ step quantum mechanical algorithm to examine an unsorted database containing $N$ items, for a single item that satisfies a particular property. The next step is to explore what statistics can be obtained by quantum mechanical algorithms, faster than is possible with classical algorithms.

Consider an experiment that can give two results with slightly different probabilities of $\left(\frac{1}{2}+\varepsilon\right)$ and $\left(\frac{1}{2}-\varepsilon\right)$ respectively where the value of $\varepsilon$ is unknown. The experiment is to be repeated several times in order to estimate $\varepsilon$. If the experiment is repeated $\nu$ times, it can be easily shown that, with a large probability, the number of occurrences of the first event will lie in the range $\nu\left(\frac{1}{2}+\varepsilon\right)\pm O(\sqrt{\nu})$. Therefore in order for the uncertainty due to $O(\sqrt{\nu})$ to become less than $\nu|\varepsilon|$, $O(\sqrt{\nu})$ must be smaller than $\nu|\varepsilon|$, equivalently $\nu$ must be greater than $\Omega\left(\frac{1}{\varepsilon^2}\right)$. Therefore in order to estimate the median with a precision of $\varepsilon$ (i.e. the number of items with values smaller than $\mu$ is less than $\frac{N}{2}(1+|\varepsilon|)$ and those with values greater than $\mu$ is also less than $\frac{N}{2}(1+|\varepsilon|)$) will need $\Omega\left(\frac{1}{\varepsilon^2}\right)$ samples [Wilks43].

**1.2 Quantum mechanical algorithms** A good starting point to think of quantum mechanical algorithms is probabilistic algorithms [BV93] (e.g. simulated annealing). In these algorithms, instead of the system being in a specified state, it is in a distribution over various states with a certain probability of being in each state. At each step, there is a certain probability of making a transition from one state to another. The evolution of the system is obtained by premultiplying this probability vector (that describes the distribution of probabilities over various states) by a state transition matrix. Knowing the initial distribution and the state transition matrix, it is possible in principle to calculate the distribution at any instant in time.

Just like classical probabilistic algorithms, quantum mechanical algorithms work with a probability distribution over various states. However, unlike classical systems, the probability vector does not completely describe the system. In order to completely describe the system we need the *amplitude* in each state which is a complex number.

The evolution of the system is obtained by premultiplying this amplitude vector (that describes the distribution of amplitudes over various states) by a transition matrix, the entries of which are complex in general. The probabilities in any state are given by the square of the absolute values of the amplitude in that state. It can be shown that in order to conserve probabilities, the state transition matrix has to be unitary [BV93].

The machinery of quantum mechanical algorithms is illustrated by discussing the three operations that are needed in the algorithm of this paper. The first is the creation of a configuration in which the amplitude of the system being in any of the $2^n$ basic states of the system is equal; the second is the Fourier transformation operation and the third the selective rotation of different states.

A basic operation in quantum computing is that of a "fair coin flip" performed on a single bit whose states are 0 and 1 [Simon94]. This operation is represented by the following matrix: $M = \frac{1}{\sqrt{2}}\begin{bmatrix} 1 & 1 \\ 1 & -1 \end{bmatrix}$. A bit in the state 0 is transformed into a superposition in the two states: $\left(\frac{1}{\sqrt{2}}, \frac{1}{\sqrt{2}}\right)$. Similarly a bit in the state 1 is transformed into $\left(\frac{1}{\sqrt{2}}, -\frac{1}{\sqrt{2}}\right)$, i.e. the magnitude of the amplitude in each state is $\frac{1}{\sqrt{2}}$ but the *phase* of the amplitude in the state 1 is inverted. The phase does not have an analog in classical probabilistic algorithms. It comes about in quantum mechanics since the amplitudes are in general complex. This results in *interference* of different possibilities as in wave mechanics and is what distinguishes quantum mechanical systems from classical systems.

In a system in which the states are described by $n$ bits (it has $2^n$ possible states) we can perform the transformation $M$ on each bit independently in sequence thus changing the state of the system. The state transition matrix representing this operation will be of dimension $2^n \times 2^n$. In case the initial configuration was the configuration with all $n$ bits in the first state, the resultant configuration will have an identical amplitude of $2^{-\frac{n}{2}}$ in each of the $2^n$ states. This is a way of creating a distribution with the same amplitude in all $2^n$ states.

Next consider the case when the starting state is another one of the $2^n$ states, i.e. a state described by an $n$ bit binary string with some 0s and some 1s. The result of performing the transformation $M$ on each bit will be a superposition of states described by all possible $n$ bit binary strings with amplitude of each state having a magnitude equal to $2^{-\frac{n}{2}}$ and sign either + or −. To deduce the sign, observe that from the definition of the matrix $M$, i.e. $M = \frac{1}{\sqrt{2}}\begin{bmatrix} 1 & 1 \\ 1 & -1 \end{bmatrix}$, the phase of the resulting configuration is changed when a bit that was previously a 1 remains a 1 after the transformation is performed. Hence if $\bar{x}$ be the $n$-bit binary string describing the starting state and $\bar{y}$ the $n$-bit binary string describing the resulting string, the sign of the amplitude of $\bar{y}$ is determined by the parity of the bitwise dot product of $\bar{x}$ and $\bar{y}$, i.e. $(-1)^{\bar{x}\cdot\bar{y}}$. This transformation is referred to as the Fourier transformation [DJ92]. This operation is one of the things that makes quantum mechanical algorithms more powerful than classical algorithms and forms the basis for most significant quantum mechanical algorithms.

The third transformation that we will need is the selective rotation of the phase of the amplitude in certain states. The transformation describing this for a 3 state system is of the form: $\begin{bmatrix} e^{i\phi_1} & 0 & 0 \\ 0 & e^{i\phi_2} & 0 \\ 0 & 0 & e^{i\phi_3} \end{bmatrix}$ where $i = \sqrt{-1}$ and $\phi_1, \phi_2, \phi_3$ are arbitrary real numbers. Note that, unlike the Fourier transformation and other state transition matrices, the probability in each state stays the same since the square of the absolute value of the amplitude in each state stays the same.

**2(a) The Abstracted Problem** Consider the problem described in the following figure: there are a number of states $S_1, S_2, ... S_N$, each of which has a value V(S) associated with it.

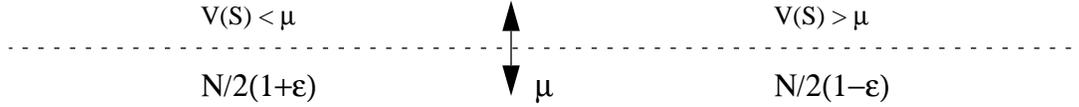

**Figure** Given a threshold $\mu$, estimate $\varepsilon$, i.e. the difference in the number of states with values above & below $\mu$.

It should be clear that if we are able to solve the problem described in the above figure, then by repeating it with different $\mu$, a logarithmic number of times, we can obtain the median. Next consider the problem described in the following paragraph in which a range for $\varepsilon$ is specified and it is required to estimate $\varepsilon$ in this range with a specified precision $\theta$. An algorithm for this problem will be presented in this paper. A little thought will show that the problem in the above figure, and hence the problem of finding the median, can be easily transformed to this problem (this is discussed further in the appendix).

Let a system contain $N = 2^n$ states which are labelled $S_1, S_2, ... S_N$. These $2^n$ states are represented as $n$ bit strings. Let each state, S, have a value, V(S), associated with it. A quantity $\mu$ is given that partitions the items such that the number of states, S, with value, V(S), less than $\mu$, is equal to $\frac{N}{2}(1+\varepsilon)$, where $|\varepsilon|$ is known to be in the range $0.1\varepsilon_0 < |\varepsilon| < \varepsilon_0$ for a specified $\varepsilon_0$ (where $\varepsilon_0$ is assumed to be smaller than 0.1). The problem is to estimate $\varepsilon$ with a precision of $\theta$ (where $\theta$ is positive) i.e. if $\varepsilon_1$ be the estimate of $\varepsilon$, then with a probability approaching 1, the true $\varepsilon$ lies in the range: $\varepsilon_1(1-\theta) < \varepsilon < \varepsilon_1(1+\theta)$. The algorithm of this paper solves the problem in $O\left(\frac{1}{\varepsilon_0 \theta^2}\right)$ time steps. The problem of finding the median is easily transformed to this, as indicated in the appendix (incidentally, even to solve this abstracted problem classically, will need a time $\Omega\left(\frac{1}{\varepsilon_0^2 \theta^2}\right)$ for the same reason indicated in section 1.1).

**2(b) Some notation** In addition to the symbols defined in section 2(a) above, the following notation will be used:

**(i)** In all steps of the algorithm, the amplitudes of all states, S, with values $V(S) < \mu$ are the same, this amplitude will be denoted by $k$; similarly in all the steps of the algorithm, the amplitudes of all states, S, with $V(S) > \mu$ are the same, this will be denoted by $l$. The amplitudes in states, S, with $V(S) < \mu$ will be referred to as the "$k$ amplitudes"; the amplitudes in states, S, with $V(S) > \mu$ will be referred to as the "$l$ amplitudes".

**(ii)** The heart of the algorithm is the loop in step 1(iv) of the algorithm in the next section. This increases the **$k$** amplitudes in each repetition of the loop. This is done in a way so that the increase is approximately proportional to $\varepsilon$. After $\nu$ repetitions of the loop, the value of the $k$ amplitude is denoted by $k_\nu(\varepsilon)$ & the value of the $l$ amplitude is denoted by $l_\nu(\varepsilon)$.

**(iii)** As mentioned in section 1.2, the probabilities are given by the square of the absolute values of the amplitudes. Hence the sum of the squares of the absolute values of the amplitudes in all states must be unity. However, in this paper we take the sum of the squares of the absolute values in the states to be $N$, where $N$ is the number of states - normalized amplitudes can be obtained by dividing the amplitudes given in this paper by $\sqrt{N}$.

**(iv) O(n) & $\Omega$(n)** (this is standard notation) A function $f(n)$ is said to be $O(n)$ if there exists a constant $\kappa$ such that for all $n$ greater than some specified $n_0$, $f(n) < \kappa n$. Similarly a function $f(n)$ is said to be $\Omega(n)$ if there exists a constant $\kappa$ such that for all $n$ greater than some specified $n_0$, $f(n) > \kappa n$.

## 3 Algorithm

This algorithm solves the problem described in 2(a): given $\mu, \varepsilon_0, \theta$, where $\varepsilon_0, \theta > 0$, it finds an estimate of $\varepsilon$ such that if $\varepsilon_1$ be this estimate, then with a high probability, the true $\varepsilon$ lies in the range: $\varepsilon_1(1-\theta) < \varepsilon < \varepsilon_1(1+\theta)$.

(1) Repeat the steps, (i)....(iv), $\alpha = O\left(\dfrac{1}{\theta^2}\right)$ times.

   (i) Initialize the system to the distribution $(1, 1, 1....1)$, i.e. there is the same amplitude to be in each of the $N$ states. This distribution can be obtained in $O(\log N)$ steps as discussed in section 1.2.

   (ii) In case the system is in a state with $V(S) > \mu$, rotate the phase by $\dfrac{\pi}{2}$ radians.

   (iii) Apply the shift transform $S$ which is defined by the $N \times N$ matrix $S$ as follows:
   $$S_{pq} = \frac{1}{N} + \frac{i}{N}, \text{ if } p \neq q; \quad S_{pp} = \frac{1}{N} - i\left(\frac{N-1}{N}\right).$$
   This shift transform, $S$, can be implemented as $S = FRF$, where $F$ is the Fourier Transform Matrix as discussed in section 1.2 and $R$ is:
   $R_{pq} = 0$ if $p \neq q$ & $R_{pp} = 1$ if $p = 0$
   & $R_{pp} = -i$ if $p \neq 0$.
   As discussed in section 1.2, the Fourier transformation matrix $F$ is defined as
   $F_{pq} = 2^{-n/2}(-1)^{\bar{p} \cdot \bar{q}}$, where $\bar{p}$ is the binary representation of $p$, and $\bar{p} \cdot \bar{q}$ is the bitwise dot product of the $n$ bit strings $\bar{p}$ & $\bar{q}$.

   (iv) Repeat the following steps - (a), (b), (c) & (d), $\beta = O\left(\dfrac{1}{\varepsilon_0}\right)$ times.

   (a) In case the system is in a state, S, with $V(S) < \mu$, rotate the phase by $\pi$ radians.

   (b) Apply the diffusion transform $D$ which is defined by the matrix $D$ as follows:
   $$D_{pq} = \frac{2}{N} \text{ if } p \neq q \quad \& \quad D_{pp} = -1 + \frac{2}{N}.$$
   This diffusion transform, $D$, can be implemented as $D = FTF$, where $F$ is the Fourier Transform Matrix, as discussed in (iii), and $T$ is defined as:
   $T_{pq} = 0$ if $p \neq q$ & $T_{pp} = 1$ if $p = 0$
   & $T_{pp} = -1$ if $p \neq 0$.

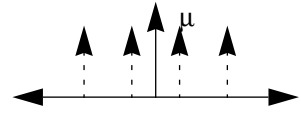

Step (1)(i): States with values above & below $\mu$, have the same amplitudes.

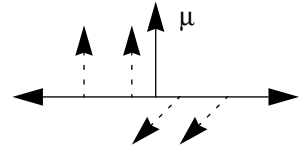

Step(1)(ii): States with values above $\mu$ have the phases of their amplitudes rotated by $90^0$. The amplitude vector becomes *(1, 1, i, i)*.

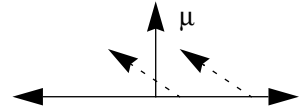

Step(1)(iii): The system shifts to states with values above $\mu$. The amplitude vector becomes *(0, 0, 1+i, 1+i)*.
*(Note that in case the number of states on the two sides is different, then there will be some amplitudes in states with $V(S) < \mu$).*

In each repetition of loop (1)(iv)(a)...(d) the $k$ amplitude increases by $\Omega(\varepsilon)$.
Therefore in $\beta$ repetitions, the $k$ amplitude becomes $\Omega(\beta\varepsilon)$.

(c) In case the system is in a state S, with
V(S) > μ, rotate the phase by π radians.
(d) Apply the diffusion transform D as described in (b) above.

(v) Sample and record the resulting state.

(2) The relative probability of the system being in a state with V(S) < μ is $\frac{1}{2}\left(|k_\beta(\varepsilon)|^2(1+\varepsilon)\right)$. In case the fraction, $f$, of the $\alpha$ samples (obtained in step 1(v) above) be calculated, then with a confidence level greater than $\left(1-2\exp\left(-2\kappa^2\right)\right)$, the quantity $\frac{1}{2}\left(|k_\beta(\varepsilon)|^2(1+\varepsilon)\right)$ will lie in the range $(f-\kappa\theta)$ to $(f+\kappa\theta)$ (theorem 8). Knowing these bounds for the function $|k_\beta(\varepsilon)|^2(1+\varepsilon)$; $\varepsilon$ can be bounded within a range of $O\left(\frac{\kappa\theta}{\beta}\right)$ with the same confidence level (theorem 9). As defined in step 1(iv), $\beta = \Omega\left(\frac{1}{\varepsilon}\right)$. It follows that the maximum expected error in $\varepsilon$ will be $O(\kappa\theta\varepsilon)$.

The $k$ amplitude (which is $\Omega(\beta\varepsilon)$) is estimated by taking $\alpha$ samples ($\alpha$ is defined in step 1).

## 4. How does the algorithm work?

Step 1(i) initializes the system so that there is an equal amplitude to be present in all of the $N = 2^n$ states. In step 1(ii), states with value greater than μ, have the phases of their amplitudes rotated by $\frac{\pi}{2}$ radians. Next, in step 1(iii), a *shift transform* is carried out by which the system *shifts* to the states with value greater than μ, i.e. the $k$ amplitudes are approximately zero and the $l$ amplitudes are of magnitude approximately $\sqrt{2}$. In case there were an equal number of states with values above and below μ, the amplitude in states with value smaller than μ will be zero. In general, if the difference in the number of states above and below μ is $N\varepsilon$, the amplitudes in states with values smaller than μ will be proportional to $\varepsilon$. This is somewhat interesting since such a redistribution could not be carried out classically. However the probabilities are proportional to the squares of the amplitudes; therefore if we try to estimate $\varepsilon$ by sampling the distribution and counting the number of samples with values below μ, it will take $\Omega\left(\frac{1}{\varepsilon^2}\right)$ samples before we obtain even a single sample with value below μ. As discussed in the introduction (section 1.1), this time is of the same order as would be required to estimate $\varepsilon$ classically.

This paper presents a technique by which the magnitudes of the $k$ amplitudes can be increased by an additive constant of $\Omega(|\varepsilon|)$ in each iteration sequence. Therefore by repeating this iteration sequence $\Omega\left(\frac{1}{\varepsilon_0}\right)$ times, the magnitudes of the $k$ amplitudes can be made to be $\Omega\left(\frac{|\varepsilon|}{\varepsilon_0}\right)$. By repeating this process $\Omega\left(\frac{1}{\theta^2}\right)$ times and observing the fraction of samples with values less than μ, $\frac{|\varepsilon|}{\varepsilon_0}$ can be estimated with a maximum error of $O(\theta)$ and thus $|\varepsilon|$ can be estimated with a maximum error of $O(\varepsilon_0 \theta)$. The heart of the algorithm is this iteration sequence for increasing the magnitudes of the $k$ amplitudes by $\Omega(|\varepsilon|)$ and by a quantity proportional to $|\varepsilon|$. This process is carried out in steps 1(iv)(a)...(d) of the following algorithm, its logic is discussed in the rest of this section.

One key step in the loop in steps 1(iv)(a)...(d) is the diffusion transform $D$. By means of theorems 4(a) & 4(b) and by using superposition, it is possible to study the effect of the diffusion transform on arbitrary values of $k$ & $l$.

**Theorem 4 (a)** - In case the initial amplitudes be $k = 1$, $l = 0$, then after applying the diffusion transform $D$, the amplitudes become $k = \varepsilon$, $l = 1 + \varepsilon$.

**Theorem 4 (b)** - In case the initial amplitudes be $k = 0$, $l = 1$, then after applying the diffusion transform $D$, the amplitudes become $k = 1 - \varepsilon$, $l = -\varepsilon$.

Using theorems 4(a) & (b), it is possible to study the evolution of arbitrary initial amplitudes by using superposition. Let the initial distribution before steps 1(iv)(a)...(d) be given by $k = K$ & $l = L$.

(a) After step 1(iv)(a), the distribution becomes $k = -K$ & $l = L$.

(b) The distribution after step 1(iv)(b) is obtained by theorems 4(a) & (b) and by superposition. Applying this gives: $k = -K\varepsilon + L(1 - \varepsilon)$ & $l = -(1 + \varepsilon)K - \varepsilon L$.

(c) The distribution after step 1(iv)(c) is obtained by changing the signs of the $l$ amplitudes. This gives:
$k = -K\varepsilon + L(1 - \varepsilon)$ & $l = (1 + \varepsilon)K + \varepsilon L$.

(d) The distribution after step 1(iv)(d) is obtained by theorems 4(a) & (b) and by superposition. Applying this gives: $k = K(1 - 2\varepsilon^2) + L(2\varepsilon - 2\varepsilon^2)$ & $l = -K(2\varepsilon + 2\varepsilon^2) + L(1 - 2\varepsilon^2)$. This is theorem 7.

**Theorem 7** - In case the initial amplitudes before step 1(iv)(a) of the algorithm be $k$ & $l$, then after steps 1(iv)(a)...(d), the new amplitudes, $k$ & $l$ are $k = K(1 - 2\varepsilon^2) + L(2\varepsilon - 2\varepsilon^2)$ and $l = K(-2\varepsilon - 2\varepsilon^2) + L(1 - 2\varepsilon^2)$ respectively.

As in [BBHT96], the evolution of the amplitudes in successive repetitions of the loop can be followed by the following corollary:

**Corollary 7.2** - For the algorithm of section 2, the $k$ amplitudes after $r$ repetitions of the loop in step 1(iv) are given by $k_r = \gamma(1 + \varepsilon + i)\sin r\phi + \varepsilon \cos r\phi$, where $\gamma^2 \equiv \dfrac{2\varepsilon - 2\varepsilon^2}{2\varepsilon + 2\varepsilon^2}$ and $\phi$ is defined by the following equation:
$\cos\phi \equiv 1 - 2\varepsilon^2$.

By definition, $\cos\phi \equiv 1 - 2\varepsilon^2$; therefore for small $\varepsilon$, $\phi \cong \sin\phi \cong 2\varepsilon$. For small values of $r\phi$, $\sin r\phi \cong r\phi \cong 2r\varepsilon$ & $\cos r\phi \cong 1$. $\gamma$, as defined in cor. 7.2 above, is approximately equal to 1 for small $\varepsilon$. Therefore by cor. 7.2, $|k_r| \cong 2\sqrt{2} r|\varepsilon|$. Therefore in $r = \Omega\left(\dfrac{1}{\varepsilon}\right)$ repetitions of the loop of 1(iv), $|k_r| \cong \Omega(1)$ and it is also proportional to $|\varepsilon|$. Thus by repeating the experiment described in step 1 of the algorithm $\Omega\left(\dfrac{1}{\theta^2}\right)$ times, it is possible to estimate $|\varepsilon|$ with a maximum relative error of $O(\theta)$.

## 5 Proofs

The following section proves that the system discussed in section 3 is indeed a valid quantum mechanical system and it gives the desired result with a probability O(1). The proofs are divided into three parts. First, it is proved that the system is a valid quantum mechanical system; second, that in each repetition of the loop of step 1(iv), the magnitudes of the $k$ amplitudes increases by $\Omega(|\varepsilon|)$; finally it is proved that by calculating the fraction of the $\alpha = O\left(\frac{1}{\theta^2}\right)$ samples with values less than $\mu$, the parameter $|\varepsilon|$ can be estimated with a precision of $\theta$.

### 5(a) Preliminary results

In order to prove that the algorithm of section 3 is a valid quantum mechanical system we need to prove that $D$ & $S$ are unitary and can be implemented as a product of local transition matrices. The diffusion transform $D$ & the shift transform $S$, are defined by the $N \times N$ matrices $D$ & $S$, as follows:

(5.0) $\quad D_{pq} = \frac{2}{N}$, if $p \neq q$ & $D_{pp} = -1 + \frac{2}{N}$.

(5.1) $\quad S_{pq} = \frac{1}{N} + \frac{i}{N}$, if $p \neq q$ & $S_{pp} = \frac{1}{N} - i\left(\frac{N-1}{N}\right)$.

Since $D$ & $S$ can be implemented as products of unitary matrices (as required in steps 1(iv)(b) and 1(iii) respectively, and as proved in theorems 2(a) and 2(b)), both $D$ & $S$ can be seen to be unitary. However, since a direct proof is short and simple, a few lines will be devoted to this. The proof for unitarity of the matrix $D$ is the same as in [Gro96].

**Theorem 1(a)** - $D$ is unitary.
**Proof** - For a matrix to be unitary it is necessary and sufficient to prove that the columns are orthonormal. The magnitude of each column vector, $q$, is $\sum_{p=1}^{n} D_{pq}^2$, $q = 1,2...n$. Substituting from (5.0), it follows that this is $\left(1 - \frac{2}{N}\right)^2 + (N-1)\frac{4}{N^2}$ which is equal to 1. The dot product of two column vectors is $\sum_{j=1}^{n} D_{jp} D_{jq}$. This is $N\frac{4}{N^2} - \frac{4}{N} = 0$.

**Theorem 1(b)** - $S$ is unitary.

**Proof** - The magnitude of each column vector, $q$, is $\sum_{p=1}^{n} S_{pq}^2$, $q = 1,2...n$. Substituting from (5.1), it follows that this is $\frac{1}{N^2} + \frac{(N-1)^2}{N^2} + 2\frac{(N-1)}{N^2}$ which is equal to 1. The dot product of two distinct column vectors is $\sum_{r=1}^{n} S_{rp} S_{rq}^*$ where $S_{rq}^*$ denotes the complex conjugate of $S_{rq}$. Evaluating $\sum_{r=1}^{n} S_{rp} S_{rq}^*$ gives:

$\left(\frac{1}{N} - i\frac{(N-1)}{N}\right)\left(\frac{1}{N} - \frac{i}{N}\right) + \left(\frac{1}{N} + \frac{i}{N}\right)\left(\frac{1}{N} + i\frac{(N-1)}{N}\right) + 2\frac{(N-2)}{N^2}$ which is equal to zero.

The way *D* & *S* are presented above, they are not local transition matrices since there are transitions from each state to all *N* states. Using the Fourier transformation matrix as defined in section 3, they can be implemented as a product of three unitary transformations, each of which is a local transition matrix.

**Theorem 2(a)** - *D* can be expressed as $D = FTF$, where *F* is the Fourier Transform Matrix as mentioned in section 1.2, and *R* is defined as follows: $T_{pq} = 0$ if $p \neq q$ & $T_{pp} = 1$ if $p = 0$ & $T_{pp} = -1$ if $p \neq 0$.

**Proof** - We evaluate *FTF* and show that it is equal to *D*. As discussed in section 3, $F_{pq} = 2^{-n/2}(-1)^{\bar{p} \cdot \bar{q}}$, where $\bar{p}$ is the binary representation of *p*, and $\bar{p} \cdot \bar{q}$ denotes the bitwise dot product of the two *n* bit strings $\bar{p}$ and $\bar{q}$. *T* can be written as $T = T_1 + T_2$ where $T_1 = -I$, *I* is the identity matrix and $T_{2,00} = 2$, $T_{2,pq} = 0$ if $p \neq 0, q \neq 0$. It is easily proved (it follows since the matrix *M*, as defined in section 1.2, is its own inverse) that $FF=I$ and hence $D_1 \equiv FT_1F = -I$. We next evaluate $D_2 \equiv FT_2F$. By standard matrix multiplication $D_{2,ad} = \sum_{bc} F_{ab} T_{2,bc} F_{cd}$.

Using the definition of $T_2$, it follows that $D_{2,ad} = 2F_{a0}F_{0d} = \frac{2}{2^n}(-1)^{\bar{a} \cdot \bar{0} + \bar{0} \cdot \bar{d}} = \frac{2}{N}$. Thus all elements of the matrix $D_2$ equal $\frac{2}{N}$. Adding $D_1$ and $D_2$ the result follows.

**Theorem 2(b)** - *S* can be expressed as $S = FRF$, where *F* is the Fourier Transform Matrix as mentioned in section 1.2, and *R* is defined as follows: $R_{pq} = 0$ if $p \neq q$ & $R_{pp} = 1$ if $p = 0$ & $R_{pp} = -i$ if $p \neq 0$.

**Proof** - We evaluate *FRF* and show that it is equal to *S*. As discussed in section 3, $F_{pq} = 2^{-n/2}(-1)^{\bar{p} \cdot \bar{q}}$, where $\bar{p}$ is the binary representation of *p*, and $\bar{p} \cdot \bar{q}$ denotes the bitwise dot product of the two *n* bit strings $\bar{p}$ and $\bar{q}$. *R* can be written as $R = R_1 + R_2$ where $R_1 = -iI$, *I* is the identity matrix (with ones on the diagonal) and $R_{2,00} = 1+i$, $R_{2,pq} = 0$ if $p \neq 0, q \neq 0$. As mentioned in theorem 2(a), $FF=I$ and hence $S_1 \equiv FR_1F = -iI$. We next evaluate $S_2 \equiv FR_2F$. By standard matrix multiplication $S_{2,ad} = \sum_{bc} F_{ab} R_{2,bc} F_{cd}$. Using the definition of $R_2$, it follows that $S_{2,ad} = (1+i)F_{a0}F_{0d} = \frac{(1+i)}{2^n}(-1)^{\bar{a} \cdot \bar{0} + \bar{0} \cdot \bar{d}} = \frac{(1+i)}{N}$. Thus all elements of the matrix $S_2$ equal $\frac{(1+i)}{N}$. Adding $S_1$ and $S_2$ the result follows.

**5 (b) Loop of step 1(iv)** Theorems 5,6 & 7 prove that the magnitudes of the amplitudes in states with values smaller than $\mu$ (i.e. the $k$ amplitudes) increase by $O(|\varepsilon|)$ as a result of each repetition of the loop in steps 1(iv)(a)...(d). Hence in $O\left(\frac{1}{|\varepsilon|}\right)$ repetitions of the loop, the amplitudes of those states will attain a value $O(1)$.

**Theorem 3** - After step 1(iii) of the algorithm, the state vector is as follows: $k = \varepsilon$, $l = (1+\varepsilon) + i$.

**Proof** - After step 1(i), the amplitudes in all states are equal, i.e. $k = 1$, $l = 1$; after step 1(ii), the $l$ amplitudes have their phase rotated by $\frac{\pi}{2}$ radians and the amplitudes become $k = 1$, $l = i$. The matrix $S$ for the shift transform is an $N \times N$ square matrix defined as follows: $S_{pq} = \frac{1}{N} + \frac{i}{N}$, if $p \neq q$ & $S_{pp} = \frac{1}{N} - i\left(\frac{N-1}{N}\right)$.

The amplitudes $k$ & $l$ after step 1(iii) thus become:

$$k = \left(\frac{1}{N} - i\left(\frac{N-1}{N}\right)\right) + \left(\frac{N}{2}(1+\varepsilon) - 1\right)\left(\frac{1+i}{N}\right) + \frac{N}{2}(1-\varepsilon)i\left(\frac{1+i}{N}\right) = \varepsilon$$

$$l = i\left(\frac{1}{N} - i\left(\frac{N-1}{N}\right)\right) + i\left(\frac{N}{2}(1-\varepsilon) - 1\right)\left(\frac{1+i}{N}\right) + \frac{N}{2}(1+\varepsilon)\left(\frac{1+i}{N}\right) = (1+\varepsilon) + i$$

**Theorem 4 (a)** - In case the initial amplitudes are $k = 1$, $l = 0$, then after applying the diffusion transform $D$, the amplitudes become $k = \varepsilon$, $l = 1 + \varepsilon$.

**Proof** - As mentioned before (4.0), the diffusion transform $D$ is defined by the $N \times N$ matrix $D$ as follows:

(5.0) $\qquad D_{pq} = \frac{2}{N}$, if $p \neq q$ & $D_{pp} = -1 + \frac{2}{N}$.

Therefore the amplitudes after applying $D$ become:

$$k = -\left(1 - \frac{2}{N}\right) + \frac{2}{N}\left(\frac{N}{2}(1+\varepsilon) - 1\right) = \varepsilon$$

$$l = \frac{2}{N} \cdot 1 \cdot \frac{N}{2} \cdot (1+\varepsilon) = (1+\varepsilon).$$

**Theorem 4 (b)** - In case the initial amplitudes are $k = 0$, $l = 1$, then after applying the diffusion transform $D$, the amplitudes become $k = 1 - \varepsilon$, $l = -\varepsilon$.

**Proof** - As in theorem 4(a) above, the amplitudes after $D$ become:

$$k = \frac{2}{N} \cdot 1 \cdot \frac{N}{2} \cdot (1-\varepsilon) = (1-\varepsilon)$$

$$l = -\left(1 - \frac{2}{N}\right) + \frac{2}{N}\left(\frac{N}{2}(1-\varepsilon) - 1\right) = -\varepsilon.$$

As is well known, in a unitary transformation the total probability is conserved - this is proved for the particular case of the diffusion transformation. This enables us to follow the evolution of the algorithm while keeping track of only one of the two quantities $k$ or $l$, not both.

**Theorem 5** - In case the initial amplitudes be $k = K$, $l = L$, then after applying the diffusion transform $D$, the new amplitudes $k = K_1$, $l = L_1$ satisfy $K_1^2(1+\varepsilon) + L_1^2(1-\varepsilon) = K^2(1+\varepsilon) + L^2(1-\varepsilon)$.

**Proof** - Using theorems 4(a) & 4(b), it follows by superposition that the new amplitudes after the diffusion transform, $D$, are given by $K_1 = K\varepsilon + L(1-\varepsilon)$ & $L_1 = K(1+\varepsilon) - L\varepsilon$. Calculating $\left(K_1^2(1+\varepsilon) + L_1^2(1-\varepsilon)\right)$, the result follows.

In order to follow the evolution of the $k$ & $l$ amplitudes due to successive repetitions of the loop in step 1(iv) (which is done in theorem 7), we first prove theorem 6:

**Theorem 6 (a)** - In case the initial amplitudes before step (a) of the algorithm be $k = 1$, $l = 0$, then after steps (a), (b), (c) & (d), the amplitudes are $k = 1 - 2\varepsilon^2$, $l = -2\varepsilon - 2\varepsilon^2$.

**Proof** - After step (a), the amplitudes become $k = -1$, $l = 0$. It follows by theorem 4(a) that after step (b) the amplitudes become $k = -\varepsilon$, $l = -(1 + \varepsilon)$; after step (c) the amplitudes become $k = -\varepsilon$, $l = (1 + \varepsilon)$. Using theorem 4(a) & 4(b), it follows by superposition that after step (d), the amplitudes are $k = -\varepsilon \cdot \varepsilon + (1 + \varepsilon) \cdot (1 - \varepsilon) = \left(1 - 2\varepsilon^2\right)$ & $l = -\varepsilon(1 + \varepsilon) - (1 + \varepsilon)\varepsilon = -2\varepsilon - 2\varepsilon^2$.

**Theorem 6 (b)** - In case the initial amplitudes before step (a) of the algorithm be $k = 0$, $l = 1$, then after steps (a), (b), (c) & (d), the amplitudes are $k = 2\varepsilon - 2\varepsilon^2$, $l = 1 - 2\varepsilon^2$.

**Proof** - After step (a), the amplitudes stay unchanged as $k = 0$, $l = 1$. It follows by theorem 4(b) that after step (b) the amplitudes become $k = 1 - \varepsilon$, $l = -\varepsilon$; after step (c) the amplitudes become $k = (1 - \varepsilon)$, $l = \varepsilon$. Using theorems 4(a) & 4(b), it follows that after step (d), the amplitudes are $k = \varepsilon \cdot (1 - \varepsilon) + \varepsilon \cdot (1 - \varepsilon) = \left(2\varepsilon - 2\varepsilon^2\right)$ & $l = (1 + \varepsilon)(1 - \varepsilon) - \varepsilon \cdot \varepsilon = 1 - 2\varepsilon^2$.

Finally using theorem 6, we can follow the evolution of the $k$ & $l$ amplitudes due to successive repetitions of the loop in step 1(iv).

**Theorem 7** - In case the initial amplitudes before step (a) of the algorithm be $K$ & $L$, then after steps (a), (b), (c) & (d), the new amplitudes, $k$ & $l$ are $k = K\left(1 - 2\varepsilon^2\right) + L\left(2\varepsilon - 2\varepsilon^2\right)$ and $l = K\left(-2\varepsilon - 2\varepsilon^2\right) + L\left(1 - 2\varepsilon^2\right)$.

**Proof** - Using theorem 6(a) & 6(b), it follows by superposition that the amplitudes after steps (a), (b), (c) & (d) are given by $k = K\left(1 - 2\varepsilon^2\right) + L\left(2\varepsilon - 2\varepsilon^2\right)$ & $l = -K\left(2\varepsilon + 2\varepsilon^2\right) + L\left(1 - 2\varepsilon^2\right)$.

As in [BBHT96], the evolution of the state vector is obtained in terms of the sin & cos functions:

**Corollary 7.1** - In case the initial amplitude be $k = \gamma \sin \tau$ and $l = \cos \tau$ where $\gamma^2 \equiv \dfrac{2\varepsilon - 2\varepsilon^2}{2\varepsilon + 2\varepsilon^2}$; then after $r$ repetitions of the loop 1(iv), $k_r = \gamma \sin(r\phi + \tau)$, $l_r = \cos(r\phi + \tau)$ where $\cos \phi \equiv 1 - 2\varepsilon^2$.

**Proof** - We prove the result by induction. By the initial conditions, the result clearly holds for $r = 0$. Assume it to be true for a particular $r$. We show that it holds for $(r + 1)$. By theorem 7: $\begin{bmatrix} k_{r+1} \\ l_{r+1} \end{bmatrix} = \begin{bmatrix} 1 - 2\varepsilon^2 & 2\varepsilon - 2\varepsilon^2 \\ -2\varepsilon - 2\varepsilon^2 & 1 - 2\varepsilon^2 \end{bmatrix} \begin{bmatrix} k_r \\ l_r \end{bmatrix}$. By the induction hypothesis, $k_r = \gamma \sin(r\phi + \tau)$, $l_r = \cos(r\phi + \tau)$. Expressing the transformation matrix in terms of $\gamma$ and $\phi$ from the definitions in the statement of the corollary, it follows that:

$\begin{bmatrix} k_{r+1} \\ l_{r+1} \end{bmatrix} = \begin{bmatrix} \cos \phi & \gamma \sin \phi \\ -\dfrac{1}{\gamma} \sin \phi & \cos \phi \end{bmatrix} \begin{bmatrix} \gamma \sin(r\phi + \tau) \\ \cos(r\phi + \tau) \end{bmatrix}$. By trigonometric identities for the sum of the sine and cosine functions:

$$\begin{bmatrix} \cos\phi & \gamma\sin\phi \\ -\frac{1}{\gamma}\sin\phi & \cos\phi \end{bmatrix} \begin{bmatrix} \gamma\sin(r\phi+\tau) \\ \cos(r\phi+\tau) \end{bmatrix} = \begin{bmatrix} \gamma\sin((r+1)\phi+\tau) \\ \cos((r+1)\phi+\tau) \end{bmatrix}. \quad \text{Hence} \quad \begin{bmatrix} k_{r+1} \\ l_{r+1} \end{bmatrix} = \begin{bmatrix} \gamma\sin((r+1)\phi+\tau) \\ \cos((r+1)\phi+\tau) \end{bmatrix}. \quad \text{Thus by}$$

induction, the corollary holds for all $r$.

**Corollary 7.2** - For the algorithm of section 2, the $k$ amplitudes after $r$ repetitions of the loop in step 1(iv) are given by $k_r = \gamma(1+\varepsilon+i)\sin r\phi + \varepsilon\cos r\phi$.

**Proof** - The initial amplitudes are $k = \varepsilon$ & $l = 1+\varepsilon+i$. The result follows by cor. 7.1 and superposition.

**Corollary 7.3** - There exists a number $\beta < \frac{1}{40\varepsilon_0}$ such that $|k_\beta| > \frac{|\varepsilon|}{20\varepsilon_0}$.

**Proof** - By definition (in cor. 7.1), $\cos\phi \equiv 1 - 2\varepsilon^2$; therefore for small $\varepsilon$, $\phi \cong \sin\phi \cong 2\varepsilon$. Also for small values of $r\phi$, $\sin r\phi \cong r\phi$ & $\cos r\phi \cong 1$; assuming small $\varepsilon$, it follows from the definition of $\gamma$ in cor. 7.1 that $\gamma \cong 1$. Therefore by cor. 7.2, $|k_r| \cong 2\sqrt{2}r|\varepsilon|$.

## 5(c) Estimating $\varepsilon$:
In case the probability of an event is estimated by carrying out $\alpha$ identical experiments, it follows by the law of large numbers, that the maximum expected error in the estimate is $O\left(\frac{1}{\sqrt{\alpha}}\right)$. By the argument before cor. 7.3, the amplitude of the system being found in states, S, with $V(S) < \mu$ is $\Omega(\beta\varepsilon)$, the probability is proportional to the square of the amplitude and is hence $\Omega(\beta^2\varepsilon^2) = \Omega\left(\frac{\varepsilon^2}{\varepsilon_0^2}\right)$; since the experiment of section 2 is repeated $\Omega\left(\frac{1}{\theta^2}\right)$ times, the maximum expected error in $\left(\frac{\varepsilon}{\varepsilon_0}\right)^2$ is $O(\theta)$; i.e. $K - O(\theta) < \left(\frac{\varepsilon}{\varepsilon_0}\right)^2 < K + O(\theta)$ where $K = O(1)$ (this is proved in theorem 8). Equivalently $\varepsilon_0(K_1 - O(\theta)) < |\varepsilon| < \varepsilon_0(K_1 + O(\theta))$ where $K_1 = O(1)$, hence $\varepsilon$ can be estimated with a precision of $O(\theta)$ as desired in section 2(a) (proved in theorem 9).

In order to prove theorem 8, we first state a version of the well known law of large numbers from probability theory:
**Lemma 8** - In an experiment, let the probability of the event $A$ be $p$, the probability of the complementary event $\bar{A}$ will be $(1-p)$. In $\eta$ repetitions of the experiment, the fraction of occurrences of $A$ lies between $p - \kappa\sqrt{\frac{1}{\eta}}$ and $p + \kappa\sqrt{\frac{1}{\eta}}$ with a probability of at least $\left(1 - 2\exp(-2\kappa^2)\right)$.
**Proof** - Omitted.

**Theorem 8** - In case $\alpha = O\left(\frac{1}{\theta^2}\right)$ samples be obtained (as described in step 1 of the algorithm) and a fraction $f$ have a value less than $\mu$, then with a confidence level greater than $\left(1 - 2\exp(-2\kappa^2)\right)$, the quantity $\frac{1}{2}\left(|k_\beta(\varepsilon)|^2(1+\varepsilon)\right)$ lies in the range $(f - \kappa\theta)$ to $(f + \kappa\theta)$.

**Proof** - By the definition of $k_\beta(\varepsilon)$, it follows that the probability of each state with $V(S) < \mu$ is $\dfrac{|k_\beta(\varepsilon)|^2}{N}$. Since there are $\dfrac{N}{2}(1+\varepsilon)$ such states, it follows that the combined probability of all these states is $\dfrac{1}{2}\left(|k_\beta(\varepsilon)|^2(1+\varepsilon)\right)$. Using lemma 8, with the event $A$ being the event corresponding to the occurrence of a sample with value smaller than $\mu$, the theorem follows.

Next we show that the value of $\varepsilon$ can be derived with a maximum expected error of $O(\theta\varepsilon)$ from the estimate of $|k_\beta(\varepsilon)|^2(1+\varepsilon)$, which was obtained in step 2 of the algorithm from the fraction of samples with values less than $\mu$. For this we prove a preliminary lemma:

**Lemma 9** - Let $a$ and $b$ be two arbitrary real numbers such that $a < b$. Let $\Delta f = f(b) - f(a)$ and the derivative $f'(x)$ at every $x$, such that $a < x < b$, is greater than $d$ (where $d > 0$). Then $b - a < \dfrac{\Delta f}{d}$.

**Proof** - By the remainder theorem, $f(b) - f(a) = f'(x)(b-a)$, for some $x$ such that $a < x < b$. Therefore $f(b) - f(a) > d(b-a)$.

**Theorem 9** - Let the function $f(\varepsilon)$ be $\dfrac{1}{2}\left(|k_\beta(\varepsilon)|^2(1+\varepsilon)\right)$. Let $|f(b) - f(a)| = 2\kappa\theta$ where $\beta a < 0.1$, $\beta b < 0.1$. Then $|b - a| < 10,000\kappa\theta\varepsilon_0$.

**Proof** - Follows by these steps:

(i) By cor. 7.2, $|k_\beta(\varepsilon)|^2 = |\gamma(1+\varepsilon)\sin\beta\phi + \varepsilon\cos\beta\phi|^2 + \gamma^2|\sin\beta\phi|^2$, $\phi \cong 2\varepsilon$, $\gamma \cong 1 - \varepsilon$.

(ii) $\dfrac{\partial}{\partial\varepsilon}\left(|k_\beta(\varepsilon)|^2(1+\varepsilon)\right) \cong \dfrac{\partial}{\partial\varepsilon}\left(2(\sin\beta\phi)^2\right) \cong \dfrac{\partial}{\partial\varepsilon}\left(8(\beta\varepsilon)^2\right) = 16\beta^2\varepsilon$.

(iii) By lemma 9, $|b - a| < \dfrac{|f(b) - f(a)|}{\min|f'(x)|}$. Therefore $|b - a| < \dfrac{2\kappa\theta}{8\beta^2\varepsilon} = \dfrac{\kappa\theta}{4\beta^2\varepsilon} = \dfrac{\kappa\theta}{4\beta^2\varepsilon_0}\left(\dfrac{\varepsilon_0^2}{\varepsilon^2}\right)\varepsilon < \dfrac{\kappa\theta}{4\left(\dfrac{1}{20}\right)^2}(10)^2\varepsilon$.

(the last step in (iv) follows by using cor. 7.2 by which $\beta > \dfrac{1}{20\varepsilon_0}$ & by the problem specification by which $\dfrac{\varepsilon_0}{\varepsilon} < 10$)

Therefore in case the value of $\varepsilon$ be estimated from the fraction of samples obtained in step 1(iv) with values less than $\mu$ (as described in step 2); then with a confidence level greater than $\left(1 - 2\exp(-2\kappa^2)\right)$, $\varepsilon$ will lie within $10,000\kappa\theta\varepsilon_0$ of the predicted value.

# 8 Appendix

The algorithm described in this paper actually measures the probability which is approximately proportional to $|\beta\varepsilon|^2$. This will give two estimates for $\varepsilon$ - one a positive range and the other a negative range. Just from this measurement, it is not possible to tell the sign of $\varepsilon$. However, the sign of $\varepsilon$ may be easily deduced by repeating the algorithm with a slightly perturbed value of the threshold $\mu$ and observing the changes in the estimates of $\varepsilon$. Denote the subroutine that returns this estimate for $\varepsilon$ (with the proper sign), given $\varepsilon_0$, $\theta$ & $\mu$ by

$\varepsilon\text{-}est\,(\mu, \varepsilon_0, \theta)$ .

The notation used in this appendix is the same as that used in the paper and discussed in section 2(b), i.e. $\mu$ is the threshold, $\varepsilon$ is the fractional difference in the number of elements with values above and below $\mu$, $\theta$ is the precision with which $\varepsilon$ needs to be estimated.

## (i) Estimating the number of elements above a general threshold

The problem discussed previously in the main paper assumes that both $\mu$ and $\varepsilon_0$ are given. In case, just the threshold $\mu$ is given but not $\varepsilon_0$ and $\theta$, the algorithm described in this paper could still be used by keeping $\theta$ fixed as a small quantity (e.g. 0.1) and initially choosing a high value of $\varepsilon_0$ (e.g. 0.1) and reducing it by a constant factor in each repetition. The value of the estimate that the subroutine $\varepsilon\text{-}est\,(\mu, \varepsilon_0, \theta)$ returns will be close to zero until the hypothesized $\varepsilon_0$ becomes of the same order as the actual $\varepsilon$. The following paragraph gives a more formal description of $\varepsilon\text{-}est\,(\mu, \varepsilon_{min})$.

Denote this subroutine that returns this estimate for $\varepsilon$, given only $\varepsilon_{min}$ & $\mu$ by $\varepsilon\text{-}est\,(\mu, \varepsilon_{min})$; this calls the previously defined subroutine $\varepsilon\text{-}est\,(\mu, \varepsilon_0, \theta)$ a logarithmic number of times.

```
ε-est (μ, ε_min)                           // Note that this subroutine has just two parameters,
{                                          // unlike the one discussed earlier, which has three.
        θ = 0.1, ε_0 = 0.1
        {                                  // It is assumed that both ε & ε_min are greater than 0.1.

                est = ε-est (μ, ε_0, θ)
                if (|est| > 0.2 ε_0)   break;
                else              ε_0 = |est| × 0.5
        }
         while (ε_0 > ε_min)
        return (est)
}
```

**(ii) Estimating the median** The median is estimated by using the routine of (i), i.e. $\varepsilon\text{-}est\,(\mu, \varepsilon_{min})$, as a subroutine in a binary search algorithm. It requires a logarithmic number of repetitions of $\varepsilon\text{-}est\,(\mu, \varepsilon_{min})$.

*median (min, max, $\varepsilon_{min}$)*
{
    *lower = min;*
    *upper = max;*
    *while (upper - lower > 1 )*
    {
        $\mu = \dfrac{upper + lower}{2}$
        $est = \varepsilon\text{-}est\,(\mu, \varepsilon_{min})$
        *if (est > 0)*    *upper* $= \mu$
        *else*            *lower* $= \mu$
    }
    *return ($\mu$)*
}

## 10 Acknowledgments
The author thanks Dan Gordon, Scott Decatur, Peter Shor & John Reif for going through earlier drafts of this paper.